\newif\ifeprint \eprinttrue                                          
\ifeprint \pdfoutput=1                                               
\documentclass[rmp,twocolumn,amsmath,floatfix,                       
        letterpaper]{revtex4}                                        
\makeatletter                                                        
\def\raggedcolumn@skip{\vskip\z@\@plus.0001fil\relax}\makeatother    
\usepackage[bookmarks=false]{hyperref}                               
\usepackage{array,orcidlink,graphicx,newtxtext,newtxmath}            
\def\eprint#1{E-print \urlalt{https://arxiv.org/abs/#1}{arXiv:#1}}   
\usepackage{breakurl}
\ifpdf\def\urlalt#1#2{\burlalt{#2}{#1}}\else\let\urlalt=\burlalt\fi  
\def\doi#1{\urlalt{https://doi.org/#1}{doi:#1}}                      
\hypersetup                                                          
{pdftitle={Geodesic intersections},pdfauthor={C. F. F. Karney}}      
\date{\today}                                                        
\else                                                                
\documentclass[Journal,letterpaper]{ascelike-new}
\usepackage{amsmath,array,graphicx,breakurl,newfloat}
\usepackage{newtxtext,newtxmath}
\usepackage[figurename=Fig.,labelfont=bf,labelsep=period]{caption}
\usepackage[colorlinks=true,citecolor=red,linkcolor=black]{hyperref}
\NameTag{Karney, \today}
\let\citep=\cite
\let\citet=\citeN

\def\doi#1{\url{https://doi.org/#1}}
\def\eprint#1{\url{https://arxiv.org/abs/#1}}
\fi                                                                  
\def\abs#1{\left|#1\right|}
\def\ave#1{\left<#1\right>}
\def\ceil#1{\lceil#1\rceil}
\newcommand{\atanx}[2]{\tan^{-1}\genfrac{}{}{1.4pt}{0}{#1}{#2}}
\renewcommand{\d}{\mathrm d}

\def\figuredir{figures}
\begin{document}
\title{Geodesic intersections}

\ifeprint                                                      
\author{Charles F. F. Karney\,\orcidlink                       
  {0000-0002-5006-5836}}                                       
\email[Email addresses: ]{charles.karney@sri.com}              
\thanks{\href                                                  
  {mailto:karney@alum.mit.edu}{karney@alum.mit.edu}.}          
\affiliation{\href{https://www.sri.com}{SRI International},    
201 Washington Rd, Princeton, NJ 08540-6449, USA}              
\else                                                          
\author[1]{Charles F. F. Karney}
\affil[1]{SRI International, 201 Washington Rd, Princeton, NJ 08540-6449, USA.
ORCID:~https://orcid.org/0000-0002-5006-5836.
Email:~charles.karney@sri.com;~karney@alum.mit.edu.}
\maketitle
\fi                                                            

\begin{abstract}
A complete treatment of the intersections of two geodesics on the
surface of an ellipsoid of revolution is given.  With a suitable metric
for the distances between intersections, bounds are placed on their
spacing.  This leads to fast and reliable algorithms for finding the
closest intersection, determining whether and where two geodesic
segments intersect, finding the next closest intersection to a given
intersection, and listing all nearby intersections.  The cases where the
two geodesics overlap are also treated.
\end{abstract}
\ifeprint \maketitle \fi 

\ifeprint \else 
\section*{Practical applications}

The intersection of lines plays a central role when performing geometric
operations on geographical objects.  Often, this is performed on a map
projection; but this has the disadvantage that the projection introduces
an inevitable distortion.  It is, therefore, preferable to compute the
intersections directly on the surface of the earth (or, more precisely,
on some ellipsoidal approximation to the earth); in this case, the lines
in question are best taken to be geodesics, the generalization of
straight lines to a curved surface.  This paper describes a fast,
reliable, and accurate method of computing the intersections of
geodesics.
\fi             

\section{Introduction}\label{intro}

In this paper, I provide a method for finding the intersection of two
geodesics on the surface of an ellipsoid of revolution.  This elaborates
on the method given by \citet[henceforth referred to as {\it
BML}]{baselga18} in two regards.  Firstly I correct their formulation of
the solution of the intersection problem for a sphere.  Secondly, I
ensure that the {\it closest} intersection is found.

The intersection problem was previously tackled by \citet{sjoeberg08};
however, he makes the problem unnecessarily complicated by combining the
solution of the geodesic problem and finding the intersection.  In
contrast, {\it BML} assume the solution of the inverse geodesic problem
is given by \citet{karney13}.  In \S8 of this latter paper, I also
provided a solution to the intersection problem using the ellipsoidal
gnomonic projection; however, although this projection is useful in
converting various geodesic problems to equivalent problems in plane
geometry, it is restricted to cases where the intersection is
sufficiently close.

The outline of the method given here is as follows.  In the first
instance, a geodesic is specified by its position (latitude and
longitude, $\phi_0$ and $\lambda_0$) and azimuth $\alpha_0$ and we
establish a criterion for the ``closest'' intersection of two such
geodesics.  I use the method of {\it BML}, with fixes given in
Appendix \ref{spherical}, to determine a ``nearby'' intersection.  In a
minority of cases, the resulting intersection is not the closest and a
search needs to be made for alternate intersections, and the closest one
is then picked.  Finally, we consider variants:
\begin{itemize}
\item
Find the intersection of two geodesic segments, specified by their
endpoints.
\item
Given a particular intersection, find the next closest intersection.
\item
List all the intersections within a certain distance.
\end{itemize}

We treat here ellipsoids of revolution with equatorial radius $a$ and
polar semi-axis $b$.  The shape of the ellipsoid is parameterized by the
flattening $f = {(a-b)}/a$, the third flattening $n = (a-b)/(a+b)$, or
the eccentricity $e = \sqrt{a^2-b^2}/a$.

The main application for this method is terrestrial ellipsoids for which
$f \approx \frac1{300}$.  However, if we restrict our attention to such
ellipsoids, we might overlook some ellipsoidal effects.  To avoid this,
we consider ellipsoids with third flattening satisfying $\abs n \le
0.12$; this includes the range of flattenings $-\frac14 \le
f \le \frac15$ which corresponds to $\frac45 \le b/a \le \frac54$.  (The
methods can be applied to Saturn with $f \approx\frac1{10}$.)  We then
expect to contend with the same effects as terrestrial ellipsoids, just
in an exaggerated form.  We intentionally do not treat more extreme
flattenings which may introduce qualitatively new behavior.

Because the range of flattenings considered exceeds the limit $\abs
f \le \frac1{50}$ of validity for the series approximations for
geodesics given in \citet{karney13}, we shall also use the formulation
for geodesics in terms of elliptic integrals described
in \citet{karney-geod2}.  When applying the methods to terrestrial
ellipsoids, the simpler series solution can be used.

\section{Posing the problem}

A geodesic $X$ on an ellipsoid can be specified by its starting position
$(\phi_{x0}, \lambda_{x0})$ and azimuth $\alpha_{x0}$.  Traveling a
signed displacement $x$ from the starting point gives a position
$(\phi_x, \lambda_x)$ and azimuth $\alpha_x$.  If we consider two such
geodesics $X$ and $Y$, then an intersection is given by the pair of
displacements $\mathbf S = [x, y]$ for which the geodesic distance
between $(\phi_x, \lambda_x)$ and $(\phi_y, \lambda_y)$ vanishes.

This investigation starts with the solution to the intersection problem
given by {\it BML}.  The method involved repeatedly solving a spherical
triangle given an edge and its two adjacent angles; this provides an
iterative method of determining {\it an} intersection.
Appendix \ref{spherical} outlines this method and includes recasting the
algorithm to ensure that it always converges and that, for the sphere,
it gives the closest intersection.  This ``basic algorithm'' maps a
tentative intersection $\mathbf S$ (which might be the origin $\mathbf
0$, where the geodesics are specified) to some nearby intersection
$\mathbf b(\mathbf S)$.

{\it BML} suggest using the equatorial radius $a$ as the equivalent
radius for the sphere.  However, a better choice for this method is $R
= \ave K^{-1/2}$, where $K$ is the Gaussian curvature and the average is
over the surface of the ellipsoid.  The Gaussian curvature is the
appropriate curvature for this application because this describes the
intrinsic properties of the surface regardless of how it is embedded in
three-dimensional space.  For any surface with the topology of a ball,
the Gauss-Bonnet theorem gives
\begin{equation}
\int K\,\d A = 4\pi,
\end{equation}
where $\d A$ is an area element and the integral is over the surface of
the ellipsoid.  Dividing this by
\begin{equation}
\int \,\d A = A,
\end{equation}
the total area, we obtain $\ave K = 4\pi/A$; this then establishes that
$A = 4\pi R^2$, i.e., that $R$ is the authalic radius.  For an ellipsoid
of revolution, we have
\begin{equation}
R = \sqrt{\frac{a^2}2 + \frac{b^2}2 \frac{\tanh^{-1}e}e}.
\end{equation}
In the remainder of this paper, we set $R = 1/\pi$ so that the
circumference of the authalic sphere is $2$.  (Alternatively, the reader
can understand that all lengths are in units of $\pi R$.)

\begin{figure}[tbp]
\begin{center}
\includegraphics[scale=0.75]{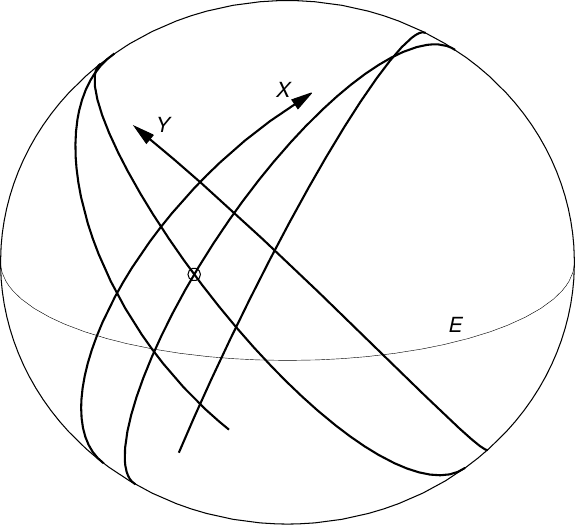}
\end{center}
\caption{\label{crossingsa}
Intersections of two geodesics $X$ and $Y$ on an ellipsoid with
flattening $f=\frac1{10}$.  This shows an orthographic view of the
geodesics on the ellipsoid; the geodesics start at a point with latitude
$\phi_{x0,y0} = 20^\circ$, longitude $\lambda_{x0,y0} = 0^\circ$ marked
with a circle with azimuths $\alpha_{x0} = 25^\circ$ and $\alpha_{y0} =
-45^\circ$ and are followed somewhat more than one complete circuit of
the ellipsoid in each direction. The equator is labeled $E$.}
\end{figure}%
Next, let us consider how to define the closest intersection.  For this
we need a measure of distance in $\mathbf S = [x, y]$ space and
plausible choices are the $L_1$~distance,
\begin{equation}\label{lone-def}
L_1(\mathbf S) = \abs{x} + \abs{y},
\end{equation}
and
the $L_\infty$~distance,
\begin{equation}\label{linf-def}
L_\infty(\mathbf S) = \max(\abs{x}, \abs{y}).
\end{equation}
(The familiar $L_2$~distance, $L_2(\mathbf S) =\sqrt{x^2 + y^2}$ makes
no sense in this context because $\mathbf S$ is not a Euclidean space.)
As we shall see, the $L_1$~distance aligns better with the way geodesics
behave and we shall adopt this way of determining closeness.  We
therefore use the simpler notation
\begin{equation}
\abs{\mathbf S} = L_1(\mathbf S).
\end{equation}
$\mathbf S$ is a vector space so the distance between two points
$\mathbf S_1 = [x_1, y_1]$ and $\mathbf S_2 = [x_2, y_2]$ is just
$\abs{\mathbf S_2 - \mathbf S_1}$.  We define $\mathbf c(\mathbf S)$ as
the intersection that is closest to $\mathbf S$; i.e., it minimizes
$\abs{\mathbf c(\mathbf S) - \mathbf S}$.

For a sphere, geodesics reduce to great circles that repeatedly
intersect at two antipodal points.  The situation is more complex on an
ellipsoid because geodesics are not, in general, closed curves, so they
repeatedly intersect at different points on the ellipsoid.  (For now, we
exclude the case of overlapping geodesics; this is discussed
\ifeprint in Sec.~\ref{coincident}.) \else 
later.)
\fi 
A typical situation is illustrated in Fig.~\ref{crossingsa} which
shows two geodesics $X$ and $Y$ that start at an intersection, marked
with a circle, on an ellipsoid with $f=\frac1{10}$.

\begin{figure}[tbp]
\begin{center}
\includegraphics[scale=0.75]{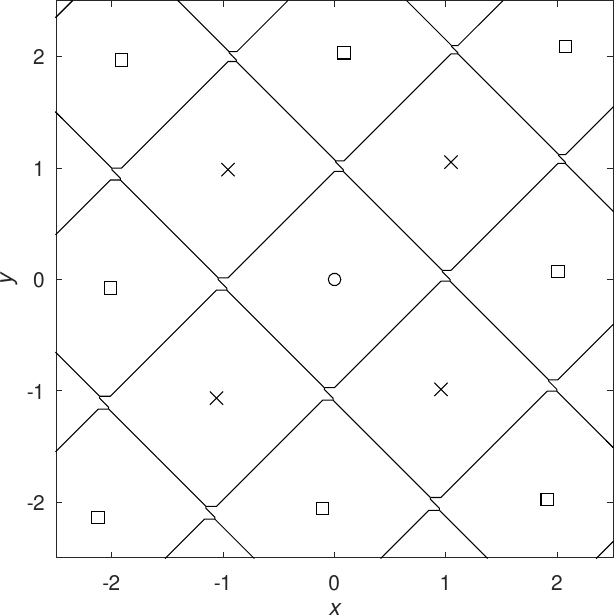}
\end{center}
\caption{\label{crossingsb}
The geodesic intersections shown in Fig.~\ref{crossingsa} laid out in
the $\mathbf S = [x, y]$ plane, where $x$ and $y$ are the displacements
along the two geodesics; the starting point is labeled with a circle;
the other 8 intersections visible in Fig.~\ref{crossingsa} are labeled
with squares; and crosses label the 4 intersections on the opposite
side of the ellipsoid.  The lines give the $L_1$~Voronoi partition.}
\end{figure}%
The intersections are also shown in the $\mathbf S = [x, y]$ plane in
Fig.~\ref{crossingsb}.  $\mathbf S$ space is partitioned into regions
(shown as lines in the figure) that are closest to the individual
intersections; this is just the Voronoi partition defined with
the $L_1$~metric.

Figure \ref{crossingsb} would look quite similar for a sphere.  In this
case, the intersections are located at integer values of $x$ and $y$
such that $x+y$ is even and the $L_1$~Voronoi partition is a regular
tiling by squares oriented at $45^\circ$.  In fact, in the case of a
sphere, the $L_2$ and $L_\infty$~partitions are identical to
the $L_1$~partition.  However, because the partitioned regions
are $L_1$~circles of radius $1$ centered at the intersections,
the $L_1$~metric yields tighter bounds on the positions of the
intersections and, as a consequence, this metric is the natural choice
for defining the closest intersection.

\begin{figure}[tbp]
\begin{center}
\includegraphics[scale=0.75]{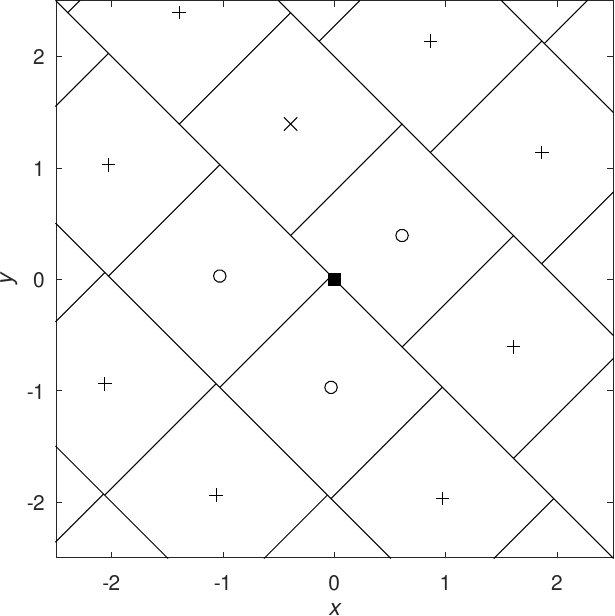}
\end{center}
\caption{\label{voronoi}
The Voronoi partition for intersections of two nearly coincident
geodesics on an ellipsoid with flattening $f=\frac1{297}$.  The
geodesics are defined by $\phi_{x0} = -50.410^\circ$, $\lambda_{x0} =
0^\circ$, $\alpha_{x0} = -69.179^\circ$, and $\phi_{y0} = 50.411^\circ$,
$\lambda_{y0} = 179.863^\circ$, $\alpha_{y0} = 68.835^\circ$.  The
starting point for the intersection problem is marked by the filled
square.}
\end{figure}%
In some cases, the pattern of intersections on an ellipsoid differs
markedly from that for a sphere.  An example is shown in
Fig.~\ref{voronoi} where the geodesics are nearly coincident.  Here, the
points defining the geodesics, namely at $\mathbf S = \mathbf 0$, are
approximately antipodal and this causes the intersection found with the
basic algorithm method $\mathbf b(\mathbf 0)$, shown by a cross in the
figure, to be far from the closest.  There are three intersections,
marked by circles, that are considerably closer.  The other
intersections are marked with plus signs.

Therefore, we seek to adapt the basic algorithm so that we are
guaranteed to find the closest intersection.  Henceforth, we will
usually drop the $L_1$ qualifier with the terms, ``closest'',
``distance'', ``radius'', and ``circle'', when used in the context of
$\mathbf S$ space.  In this paper, the boldface notation for vectors,
e.g., $\mathbf S$ and $\mathbf c(\mathbf S)$ is reserved for
two-dimensional vectors in the $[x,y]$ plane.

\section{Bounds on intersections}

A necessary prelude to constructing the solution for the closest
intersection is to place limits on intersections.  Initially, these
limits were found empirically by randomly sampling many (at least
$10^6$) cases for several values of $n$ satisfying $\abs{n} \le 0.12$.
For each value of $n$, we can identify the geodesics involved in the
limiting cases enabling an accurate determination of the limits.

Let's start by determining the minimum distance $2t_1$ between
intersections.  We find that, in the case of an oblate ellipsoid, the
minimum occurs with an intersection given by the nearly equatorial
geodesics $\phi_{x0,y0} = 0^\circ$, $\lambda_{x0,y0} = 0^\circ$,
$\alpha_{x0,y0} = 90^\circ \pm \epsilon$, for some small $\epsilon$.  In
the limit $\epsilon \rightarrow 0$, the nearest intersections are at $x
= y = \pm \pi b$, i.e., we have $\abs{[x, y]} = 2\pi b$ and hence $t_1
= \pi b$.  This result is expected, because, for an oblate ellipsoid,
the maximum curvature occurs at the equator.

This case also illustrates a common trait in these limiting cases,
namely that the geodesics are nearly coincident.  We postpone a full
discussion of such geodesics to
\ifeprint Sec.~\ref{coincident}. \else 
a later section.
\fi 
However, it's possible to find the intersections in these cases by
determining where the reduced length $m$ or geodesic scale $M$ of a
geodesic is zero.  These quantities characterize the behavior of nearby
geodesics.  Consider a geodesic starting at point 1 and ending at point
2, and a second nearby geodesic starting at the same point but a
slightly difference azimuth; the two geodesics intersect at points 2
where $m_{12} = 0$; in these cases, points 1 and 2 are called
``conjugate points''.  Furthermore, the geodesics are parallel at points
2 where $M_{21} = 0$.  Further details are given
in \citet[\S3]{karney13}; these include expressions for the derivatives
of $m_{12}$ and $M_{12}$ which allow the intersections of nearly
coincident geodesics to be found by Newton's method.

The equivalent result for $t_1$ for a prolate ellipsoid is found by
considering a geodesic passing over the pole (where the curvature is
maximum).  The geodesic, which we denote by $p_a$ should be
symmetrically placed over the pole so that $M_{21} = 0$ at the pole.
For $n = -0.01$, we find that $\phi_{x0,y0} = \phi_{x,y} = 1.755^\circ$,
$\lambda_{x0,y0} = 0^\circ$, $\lambda_{x,y} = 180^\circ$,
$\alpha_{x0,y0} = \pm\epsilon$, and $t_1 = p_a = 0.9832$.

The result for the minimum distance between intersections can be
summarized by
\begin{equation}\label{mindista}
\abs{\mathbf T' - \mathbf T} \ge 2 t_1,
\end{equation}
where $\mathbf T$ and $\mathbf T'$ are any two distinct intersections.
This result has an important corollary: if $\mathbf T$ is an
intersection satisfying $\abs{\mathbf T} < t_1$ then it is the closest
intersection to $\mathbf 0$, i.e., $\mathbf T = \mathbf c(\mathbf 0)$.
This is proved by considering the distance to some other intersection
$\abs{\mathbf T'}$ which is constrained by the triangle inequality,
\begin{equation}
\abs{\mathbf T'} + \abs{\mathbf T} \ge
\abs{\mathbf T' - \mathbf T} \ge 2t_1,
\end{equation}
or
\begin{equation}
\abs{\mathbf T'} \ge 2t_1 -  \abs{\mathbf T} \ge t_1.
\end{equation}
Thus $\mathbf T$ is closer than $\mathbf T'$, provided that
$\abs{\mathbf T} < t_1$.

We next place an upper limit on the distance to the closest
intersection.  Let $\mathbf T$ be the intersection closest to $\mathbf
0$, then $\abs{\mathbf T} \le t_2$.  The empirical results indicate that
$t_2$ is given by switching the results for $t_1$ for oblate and prolate
ellipsoids giving $t_2 = p_a$ for oblate ellipsoids and $t_2 = \pi b$
for prolate ellipsoids.  Again this result makes sense since the
corresponding geodesics are now sampling regions of minimum curvature on
the ellipsoid.

The result for the nearest intersection to a given intersection is
\begin{equation}\label{maxdista}
\abs{\mathbf T' - \mathbf T} \le 2 t_3.
\end{equation}
For prolate ellipsoids, we find that $t_3 = 2Q$, where $Q$ is the length
of the quarter meridian; $2t_3$ is the $L_1$~distance between an
intersection of two meridians at one pole and the intersection at the
opposite pole.  The oblate ellipsoids, the limiting geodesics are
oblique (labeled $o_c$) and are given by an intersection on the equator
with two nearby azimuths approximately equal to $45^\circ$ with the
azimuth adjusted so that there are 6 equally distant nearest
intersections.  For $n = 0.01$, this is achieved with
$\phi_{x0,y0}=0^\circ$, $\lambda_{x0,y0} = 0^\circ$,
$\alpha_{x0,y0}=45.062^\circ\pm\epsilon$, with nearest intersections at
$[\pm o_c, \pm o_c]$, $[\pm 1.4976, \mp 0.5058]$, and $[\pm 0.5058, \mp
1.4976]$, where $t_3 = o_c = 1.0017 = \frac12(1.4976 + 0.5058)$.

Next, we determine the ``capture radius'' $t_4$ of the basic
algorithm. If $\mathbf T$ is an intersection that satisfies
\begin{equation}
\abs{\mathbf T - \mathbf S} < t_4,
\end{equation}
where $\mathbf S$ is an arbitrary point, then
\begin{equation}
\mathbf T = \mathbf b(\mathbf S) = \mathbf c(\mathbf S).
\end{equation}
In other words, the basic algorithm applied at $\mathbf S$ is guaranteed
to return the closest intersection provided it is within a radius $t_4$
of $\mathbf S$.

\begin{table}[tbp]
\caption{\label{paramsa}
The special lengths for the intersection problem of a sphere and oblate
ellipsoids, $n \ge 0$; the authalic radius is taken to be $R = 1/\pi$.
The quantities $a$ and $b$ are the equatorial radius and the polar
semi-axis; $Q$ is the quarter meridian distance.  The other quantities
$p_a$, $p_b$, $o_c$, and $t_1$ thru $t_5$ are defined in the text.
}
\begin{center}
\begin{tabular}{@{\extracolsep{0.5em}}
      >{$}c<{$} >{$}c<{$}>{$}c<{$}>{$}c<{$}>{$}c<{$}>{$}c<{$}>{$}c<{$}}
    \hline\hline\noalign{\smallskip}
    n &  \pi a &  \pi b &     2Q &    p_a &    p_b &     o_c \\
      &        &t_1,t_4 &    t_5 &    t_2 &        &     t_3 \\
\noalign{\smallskip}\hline\noalign{\smallskip}
  0   & 1      & 1      & 1      & 1      & 1      & 1      \\
 0.01 & 1.0067 & 0.9867 & 0.9967 & 1.0165 & 1.0230 & 1.0017 \\
 0.02 & 1.0133 & 0.9735 & 0.9935 & 1.0328 & 1.0463 & 1.0034 \\
 0.03 & 1.0199 & 0.9605 & 0.9904 & 1.0487 & 1.0698 & 1.0052 \\
 0.04 & 1.0264 & 0.9475 & 0.9874 & 1.0644 & 1.0935 & 1.0071 \\
 0.05 & 1.0330 & 0.9346 & 0.9844 & 1.0796 & 1.1175 & 1.0090 \\
 0.06 & 1.0395 & 0.9218 & 0.9815 & 1.0944 & 1.1417 & 1.0110 \\
 0.07 & 1.0460 & 0.9091 & 0.9788 & 1.1088 & 1.1661 & 1.0131 \\
 0.08 & 1.0524 & 0.8965 & 0.9760 & 1.1227 & 1.1906 & 1.0153 \\
 0.09 & 1.0589 & 0.8840 & 0.9734 & 1.1361 & 1.2153 & 1.0175 \\
 0.10 & 1.0652 & 0.8716 & 0.9708 & 1.1490 & 1.2402 & 1.0198 \\
 0.11 & 1.0716 & 0.8592 & 0.9683 & 1.1613 & 1.2651 & 1.0222 \\
 0.12 & 1.0779 & 0.8469 & 0.9659 & 1.1732 & 1.2901 & 1.0246 \\
\noalign{\smallskip}
\hline\hline
\end{tabular}
\end{center}
\end{table}%
\begin{table}[tbp]
\caption{\label{paramsb}
The parameters listed in Table \ref{paramsa} computed for prolate
ellipsoids, $n < 0$.  The first row of headings match those in
Table \ref{paramsa}; however the pairing of these with $t_1$ thru
$t_5$ listed in the second row of headings
is different.}
\begin{center}
\begin{tabular}{@{\extracolsep{0.5em}}
      >{$}c<{$} >{$}c<{$}>{$}c<{$}>{$}c<{$}>{$}c<{$}>{$}c<{$}>{$}c<{$}}
    \hline\hline\noalign{\smallskip}
    n &  \pi a &  \pi b &     2Q &    p_a &    p_b &     o_c \\
      &        &    t_2 &t_5,t_3 &    t_1 &    t_4 &         \\
\noalign{\smallskip}\hline\noalign{\smallskip}
-0.01 & 0.9933 & 1.0134 & 1.0034 & 0.9832 & 0.9773 & 0.9984 \\
-0.02 & 0.9866 & 1.0269 & 1.0068 & 0.9662 & 0.9549 & 0.9967 \\
-0.03 & 0.9799 & 1.0405 & 1.0104 & 0.9491 & 0.9327 & 0.9952 \\
-0.04 & 0.9731 & 1.0542 & 1.0141 & 0.9317 & 0.9109 & 0.9936 \\
-0.05 & 0.9663 & 1.0681 & 1.0178 & 0.9143 & 0.8893 & 0.9920 \\
-0.06 & 0.9595 & 1.0820 & 1.0217 & 0.8968 & 0.8681 & 0.9905 \\
-0.07 & 0.9527 & 1.0961 & 1.0257 & 0.8792 & 0.8472 & 0.9889 \\
-0.08 & 0.9459 & 1.1103 & 1.0297 & 0.8616 & 0.8266 & 0.9873 \\
-0.09 & 0.9390 & 1.1247 & 1.0339 & 0.8440 & 0.8064 & 0.9856 \\
-0.10 & 0.9321 & 1.1392 & 1.0382 & 0.8265 & 0.7864 & 0.9839 \\
-0.11 & 0.9251 & 1.1538 & 1.0426 & 0.8089 & 0.7668 & 0.9822 \\
-0.12 & 0.9182 & 1.1686 & 1.0472 & 0.7914 & 0.7475 & 0.9803 \\
\noalign{\smallskip}
\hline\hline
\end{tabular}
\end{center}
\end{table}%
The basic algorithm switches between neighboring intersections when
$\mu_y - \mu_x = 0$ or $\pi$ in Fig.~\ref{triangle}.  For $z$ small,
this is the condition that the geodesics are parallel, and this, in
turn, is given by $M_{21} = 0$ for a geodesic starting at an
intersection.  As in the determination of $t_1$, the limiting case is
given when the geodesics lie in the region of maximum curvature.  For an
oblate ellipsoid, we have $t_4 = t_1 = \pi b$.  For a prolate ellipsoid,
it is again a geodesic passing over the pole, but this time
asymmetrically; we denote this geodesic by $p_b$.  For $n = -0.01$, we
start at an intersection $\phi_{x0,y0} = 26.783^\circ$, $\lambda_{x0,y0}
= 0^\circ$, $\alpha_{x0,y0} = \pm \epsilon$ and the condition $M_{21} =
0$ is obtained at $x = y = \frac12 p_b$, $\phi_{x,y} = 64.192^\circ$,
and $\lambda_{x,y} = 180^\circ$.  The distance from this point to the
intersection is $t_4 = p_b = 0.9773$.

When finding the intersection of geodesic segments, we require the
maximum length of a shortest geodesic segment, $t_5$.  This is just
the distance between the poles, twice the quarter meridian distance,
i.e., $t_5 = 2Q$.

These lengths are tabulated in Table \ref{paramsa} for oblate ellipsoids
with $0\le n \le 0.12$ and Table \ref{paramsb} for prolate ellipsoids
with $-0.12 \le n < 0$.  The first row of headings reflects where on the
ellipsoid the geodesics lie.  In contrast, the second row lists the
relationship with the distances $t_1$ thru $t_5$.  The ratios of the
distances $a$, $b$, $R$, $Q$ can be found by elementary means (for $Q$,
this entails solving the inverse geodesic problem between a pole and a
point on the equator).  The distances $p_a$, $p_b$, and $o_c$ are found
using iterative methods; code for these calculations is provided in the
constructor for the {\tt Intersect} class in the implementation
described
\ifeprint in Sec.~\ref{implementation}. \else 
below.
\fi                                           
However, the algorithms do not depend on the exact values of $t_1$ thru
$t_5$.  The algorithms given below will ``work'' (albeit slightly less
efficiently) if the quantities are replaced by their values at a larger
value of $\abs n$.

\section{Finding the closest intersection}

We are now in a position to consider the problem of finding the closest
intersection.  The lengths $t_1$ thru $t_4$ found in the previous
section are the radii of circles.  For example, the definition of $t_2$
can be reformulated by saying that the closest intersection is inside a
circle of radius $t_2$.  Now $L_1$~circles have the property that a
circle of radius $s$ can be exactly tiled by $m^2$ circles of radius
$s/m$.  Thus the basic algorithm can find {\it all} the intersections in
the circle of radius $t_2$ by running it with four starting points
$\{{[\pm d_1, 0]},\allowbreak {[0, \pm d_1]}\}$, where $d_1 = \frac12
t_2$.  For $\abs n \le 0.12$, we then have $d_1 < t_4$ and each
application of the basic algorithm is guaranteed to find any
intersections within a quarter of the circle of radius $t_2$;
furthermore, there is at least one intersection within this circle.  All
that remains is to pick the intersection closest to $\mathbf 0$.

We can optimize this procedure by observing that for any intersection
$\mathbf T$:
\begin{itemize}
\item
If $\abs{\mathbf T} < t_1$, then $\mathbf T$ is the intersection closest
to $\mathbf 0$, $\mathbf c(\mathbf 0)$.
\item
Otherwise, the closest intersection is either $\mathbf T$ or is an
intersection $\mathbf T'$ where $\abs{\mathbf T' - \mathbf T} \ge 2t_1$.
\end{itemize}
The first of these suggests that we start with an application of
$\mathbf b(\mathbf 0)$, and the second lets us remove certain of the
remaining starting points as the method progresses.

The algorithm for the closest intersection $\mathbf c(\mathbf 0)$ is
then
\begin{enumerate}
\item
Set $\mathbf T = [\infty, \infty]$, $d_1 = \frac12t_2$, $\delta
= \sqrt[5]\epsilon$ ($\epsilon = 2^{-52}$ for double-precision
arithmetic), $m = 5$, and $\mathbf S_i$ for $0\le i < m$ to the set of
$m$ starting points, $\{\mathbf 0,\allowbreak {[\pm d_1, 0]},\allowbreak
{[0, \pm d_1]}\}$.  Also set $e_i = \mathrm{false}$ for $0 \le i < m$;
these are ``exclusion flags'' with $e_i = \mathrm{true}$ indicating that
the starting point $\mathbf S_i$ can be skipped.
\item
For each $i$ satisfying $0\le i < m$ and $\neg e_i$ do:
\begin{enumerate}
\item
Apply the basic algorithm method starting at $\mathbf S_i$, setting
$\mathbf T_i = \mathbf b(\mathbf S_i)$.
\item
If $\abs{\mathbf T_i} < t_1$, set $\mathbf T = \mathbf T_i$ and
go to step 3.
\item
If $\abs{\mathbf T_i} < \abs{\mathbf T}$, set $\mathbf
T = \mathbf T_i$.
\item
For each $j$ satisfying $i < j < m$ and $\neg e_j$, set $e_j
= (\abs{\mathbf T_i - \mathbf S_j} < 2t_1 - d_1 - \delta)$.
\end{enumerate}
\item
Return $\mathbf T$ as the closest intersection, $\mathbf c(\mathbf 0)$.
\end{enumerate}
\begin{figure}[tb]
\begin{center}
\includegraphics[scale=0.75]{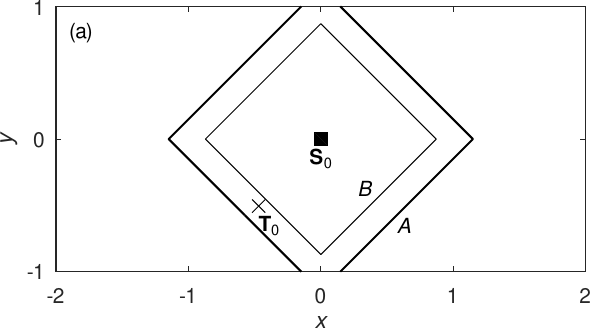}\\[1.5ex]
\includegraphics[scale=0.75]{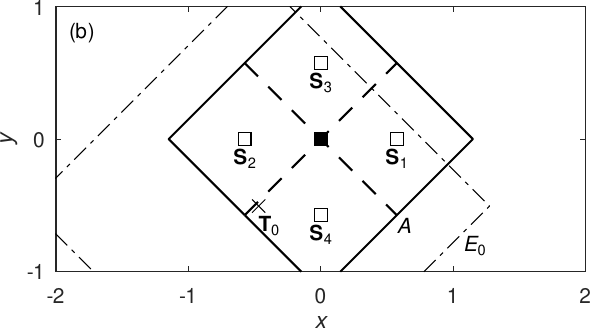}\\[1.5ex]
\includegraphics[scale=0.75]{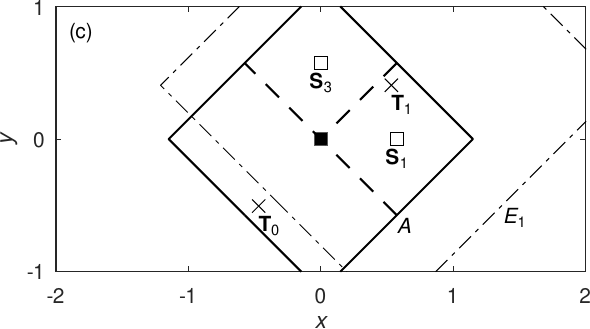}
\end{center}
\caption{\label{solve2-ex}
An example of finding the closest intersection.  Here we take $n
= \frac1{10}$, $\phi_{x0} = 36.873^\circ$, $\lambda_{x0} = 0^\circ$,
$\alpha_{x0} = 60.641^\circ$, and $\phi_{y0} = -62.631^\circ$,
$\lambda_{y0} = 75.301^\circ$, $\alpha_{y0} = 30.776^\circ$.  Parts (a),
(b), and (c) illustrate successive stages in the algorithm, as explained
in the text.}
\end{figure}%
Figure \ref{solve2-ex} illustrates this algorithm in action.
Figure \ref{solve2-ex}(a) shows the first application of the basic
algorithm, Step 2(a) for $i = 0$, gives the intersection $\mathbf T_0$.
$A$ labels the circle of radius $t_2$ inside which the closest
intersection must lie, and $B$ labels the circle of radius $t_1$ inside
which any intersection is guaranteed to be the closest.  $\mathbf T_0$
lies inside $A$ and outside $B$ (the test in Step 2(b) fails) and so all
we know is that it {\it may} be the closest intersection.  The algorithm
then repeats the basic algorithm at points $\mathbf S_i$ for $i = 1$
thru $4$.  Each application is responsible for finding intersections in
a circle of radius $d_1 = \frac12t_2$ centered at $\mathbf S_i$; this
divides the circle $A$ into 4 equal pieces delineated by dashed lines;
see Fig.~\ref{solve2-ex}(b). However, the presence of the intersection
at $\mathbf T_0$ means that there can be no intersection within the
``exclusion'' circle of radius $2t_1$ centered at $\mathbf T_0$, labeled
$E_0$ and shown as a dot-dashed line; because this encompasses the
circles allotted to $\mathbf S_2$ and $\mathbf S_4$, we can skip the
application of the basic algorithm for $i = 2$ and $4$; this is
accomplished by Step 2(d) for $i = 0$.  The second application of the
basic algorithm, $i = 1$, finds $\mathbf T_1$ which is closer than
$\mathbf T_0$, Step 2(c); see Fig.~\ref{solve2-ex}(c).  The exclusion
circle (labeled $E_1$) centered at $\mathbf T_1$ encompasses the circle
allotted to $\mathbf S_3$.  We therefore can skip the remaining
iterations and the algorithm terminates with $\mathbf T_1$ as the
closest intersection.

Even though an application of the basic algorithm at $\mathbf S_i$ can
capture intersections within a circle of radius $t_4$, we nevertheless
assign it ``responsibility'' for a {\it smaller} circle of radius $d_1$.
This allows more iterations to be skipped on account of Step 2(d).
Finally, the test in Step 2(d) includes a ``safety margin'' $\delta$;
this ensures that the exclusions don't result in missing intersections
lying {\it on} the boundaries of the circles of radius $d_1$ in
Fig.~\ref{solve2-ex}(b).

\section{The intersection of two geodesic segments}

A common application for intersections is finding if and where two
geodesic segments $X$ and $Y$ intersect.  Each segment is specified by
its endpoints, e.g., $(\phi_{x1}, \lambda_{x1})$ and
$(\phi_{x2}, \lambda_{x2})$ and we stipulate that there's a unique
shortest geodesic connecting each pair of endpoints.  As a consequence,
the two segments intersect at most once.  Displacements along the
geodesic are measured from the first endpoint, positive towards the
second endpoint. Let $s_x$ and $s_y$ be the lengths of the segments;
then the segments intersect provided that the intersection is within the
$s_x \times s_y$ rectangle $H$ specified by $0\le x \le s_x$ and $0\le
y \le s_y$.

We start by applying the algorithm for the closest intersection to the
midpoint $\mathbf M = \frac12[s_x, s_y]$, i.e., we compute $\mathbf T_0
= \mathbf c(\mathbf M)$.  If $\mathbf T_0$ is within $H$, then the
segments intersect at $\mathbf T_0$ and we're done.

Otherwise, we need to check whether there's some other intersection
within $H$.  Now $\mathbf T_0$ is the closest intersection to $\mathbf
M$.  But there may be a more distant intersection lying near the corners
of $H$; this requires
\begin{equation}\label{cornera}
\textstyle\frac12(s_x + s_y) \ge \abs{\mathbf T_0 - \mathbf M}.
\end{equation}
\begin{figure}[tb]
\begin{center}
\includegraphics[scale=0.75]{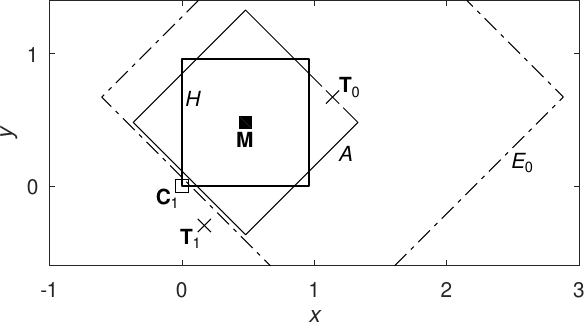}
\end{center}
\caption{\label{solve4}
An illustration of segment intersection with $n = \frac1{10}$,
$\phi_{x1} = -56.739^\circ$, $\lambda_{x1} = 0^\circ$, $\phi_{x2} =
54.809^\circ$, $\lambda_{x2} = -175.812^\circ$, and $\phi_{y1} =
-33.312^\circ$, $\lambda_{y1} = -67.388^\circ$, $\phi_{y2} =
34.788^\circ$, $\lambda_{y2} = 117.255^\circ$.}
\end{figure}%
If this inequality holds, consider each corner $\mathbf C_i$ ($0 < i \le
4$) of the intersection rectangle, $\{[0,0],\allowbreak [s_x,
0],\allowbreak [0, s_y],\allowbreak [s_x, s_y]\}$; if $\mathbf C_i$ lies
outside the exclusion radius for $\mathbf T_0$, i.e., if
\begin{equation}\label{segment-excl}
\abs{\mathbf T_0 - \mathbf C_i} \ge 2t_1,
\end{equation}
then we apply the basic algorithm and compute $\mathbf T_i = \mathbf
b(\mathbf C_i)$; if $\mathbf T_i$ lies within $H$, we accept $\mathbf
T_i$ as the result.  The procedure is illustrated in Fig.~\ref{solve4},
where the segment rectangle $H$ is shown with a heavy line.  First, the
intersection closest to the midpoint $\mathbf M$, the solid square, is
found giving $\mathbf T_0 = \mathbf c(\mathbf M)$.  Because the corners
of $H$ are further from $\mathbf M$ than $\mathbf T_0$ (outside the
circle $A$), we check for intersections close to the corners.  Only one
corner $\mathbf C_1$, marked with an open square, lies outside the
exclusion circle $E_0$ of radius $2t_1$ centered at $\mathbf T_0$, shown
with a dot-dashed line.  An application of the basic algorithm yields
the intersection $\mathbf T_1 = \mathbf b(\mathbf C_1)$.  Since this
is outside $H$, the segments do not intersect (and $\mathbf T_0$ is
returned as the closest intersection for the segments).

It turns out that we hardly ever have to compute $\mathbf b(\mathbf
C_i)$.  Consider the triangle with vertices $\mathbf M$, $\mathbf T_0$,
and $\mathbf C_i$ (for some $i$).  The triangle inequality states that
\begin{equation}\label{triangle-ineq}
\abs{\mathbf C_i - \mathbf M} \ge
\abs{\mathbf T_0 - \mathbf C_i} - \abs{\mathbf T_0 - \mathbf M}.
\end{equation}
Rewriting Eq.~(\ref{cornera}) as
\begin{equation}
\abs{\mathbf C_i - \mathbf M} \ge \abs{\mathbf T_0 - \mathbf M},
\end{equation}
and adding this to Eq.~(\ref{triangle-ineq}) gives
\begin{equation}
2\abs{\mathbf C_i - \mathbf M} \ge
\abs{\mathbf T_0 - \mathbf C_i}.
\end{equation}
Finally substituting Eq.~(\ref{segment-excl}) and $\abs{\mathbf C_i
- \mathbf M} = \frac12(s_x + s_y)$, we obtain
\begin{equation}
t_1 \le \textstyle\frac12(s_x + s_y) < t_5,
\end{equation}
where the second inequality is just the condition that $X$ and $Y$ are
unique shortest geodesics.  This represents a fairly narrow range (see
Tables \ref{paramsa} and \ref{paramsb}) which is satisfied only if the
endpoints of both $X$ and $Y$ are nearly antipodal.  The basic algorithm
finds any new intersection near a corner provided that $t_4 > t_5 -
t_1$, a condition that holds for $\abs n \le 0.12$.

For uniformly distributed endpoints, the probability that a corner needs
to be checked is roughly $1.5 \times f^5$ for small positive $f$ (this
is found empirically).  Thus the cost of finding the intersection of two
segments is only slightly greater than the cost of finding the closest
intersection.

A more consequential observation is that I have found {\it no} cases
where $\mathbf T_i$ for $i > 0$ yielded a segment intersection that
overrode $\mathbf T_0$.  So I conjecture that $\mathbf c(\mathbf M)$
{\it always} yields the intersection of two segments, provided that they
do intersect.  If this is true, then the segment intersection algorithm
just reduces to the algorithm for the closest intersection.

\section{Other intersection problems}

\begin{figure}[tb]
\begin{center}
\includegraphics[scale=0.75]{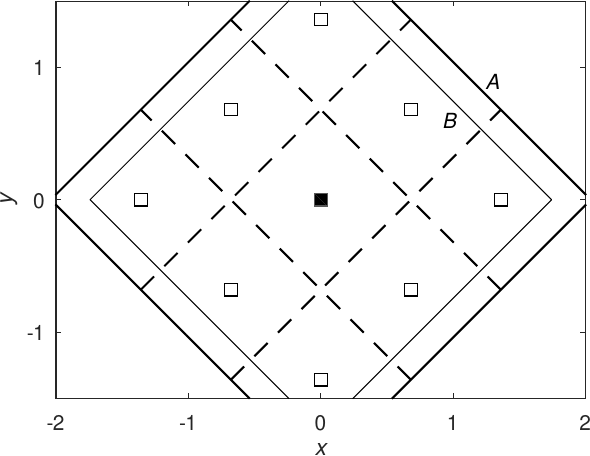}
\end{center}
\caption{\label{solve3-ex}
Finding the next closest intersection to the intersection marked with
the filled square.  Here, we have $n = \frac1{10}$.}
\end{figure}%
Our framework can be used to tackle two other intersection problems:
finding the next closest intersection to a given intersection and
enumerating all the intersections within a certain radius of a starting
point.

For the first of these problems, we know the intersection must lie
between two concentric circles of radii $2t_3$ and $2t_1$ labeled $A$
and $B$ in Fig.~\ref{solve3-ex}.  The annular region between these
circles can be covered by 8 circles of radius $d_2 = \frac23t_3$ (shown
with dashed lines), so applications of the basic algorithm starting at
the open squares in Fig.~\ref{solve3-ex} will find the next closest
intersection provided that $d_2 < t_4$, a condition that holds for $\abs
n \le 0.12$.  The modifications of the algorithm for the closest
intersection given above are: We initialize $m = 8$ and $\mathbf S_i$ is
the set of $m$ points $\{{[\pm 2d_2, 0]},\allowbreak {[0, \pm
2d_2]},\allowbreak {[\pm d_2, \pm d_2]}\}$.  Step 2(b) becomes ``If
$\mathbf T_i = \mathbf 0$, skip to the next iteration of $i$.''  For the
test in Step 2(d), replace $d_1$ with $d_2$.

Three other points: Any tests for equality between intersection points,
e.g., $\mathbf T' = \mathbf T$ should be implemented as $\abs{\mathbf T'
- \mathbf T} < \delta$; remember that intersections can't be arbitrarily
close, so if the distance between two intersections is less than
$\delta$ the discrepancy is due to roundoff error.  A minimum of 8
starting points for the basic algorithm are needed for this particular
problem because, in the spherical limit, there are 8 equidistant
intersections closest to a given intersection.  For terrestrial
ellipsoids, nearly equidistant next-closest intersections are reasonably
common---so it may be important to check other intersections; this
serves as an introduction to the next topic\ldots

The last problem we consider is listing all the intersections within
some distance $D$ from a starting point.  Given the knowledge we've
already acquired, the solution is straightforward.  Divide the circle of
radius $D$ evenly into $m^2$ circles of radius $D/m$ where $m
= \ceil{D/d_3}$ and $d_3 = t_4 - \delta$, the capture radius for the
basic algorithm reduced by the safety margin.  Run the basic algorithm
starting at the center of each of the smaller circles and gather up all
the distinct intersections lying within $D$.  (We treat intersections
closer than $\delta$ as being the same; see the previous paragraph.)
The optimization in step 2(d) in the algorithm for the closest
intersection (replacing $d_1$ by $d_3$) should be applied.

\section{Coincident geodesics}\label{coincident}

The final topic we consider is coincident geodesics.  We touched on this
in the specification of the basic algorithm in Appendix~\ref{spherical}.
This returns a flag $c$ which is normally $0$ but is set to $\pm1$ if
the geodesics are coincident at the intersection with the sign of $c$
indicating whether the geodesics are parallel or antiparallel.

Let's consider the application of the basic algorithm that returns an
intersection, $\mathbf T = [x, y] =\mathbf b(\mathbf S)$, and sets the
coincidence flag $c=\pm1$; then, for arbitrary $s$, $\mathbf T' = [x',
y'] = \mathbf T + [s,\pm s]$ is also an intersection.  When finding the
closest intersection, we are interested in the intersection that
minimizes the distance from $\mathbf 0$, i.e., which minimizes
$\abs{\mathbf T'}$.  It's easy to show graphically that
$\min(\abs{\mathbf T'}) = \abs{x \mp y}$ and that an interval of $s$ of
width $\abs{x \mp y}$ yields this minimum value.  Arguably, we could
return any value of $\mathbf T'$ which gives this value.  Nevertheless,
it is preferable to pick a definite and predictable value and for this
purpose, we pick $s = -\frac12(x \pm y)$, for which we have $\mathbf T'
= \frac12[x\mp y, y\mp x]$ .  This is the value of $s$ that also
minimizes the $L_\infty$~distance of $\mathbf T'$ from $\mathbf 0$.

A similar exercise can be carried out for the algorithm for the
intersection of segments that coincide.  In the cases where the segments
overlap, we adjust $s$ to return the intersection in the middle of the
overlap region.  Otherwise, we return the intersection in the middle of
the smallest gap between the segments.

In the case of finding the nearest intersection to a given intersection,
we distinguish two cases of coincident intersections
\begin{enumerate}
\item
An intersection $\mathbf T = [x,y]$ is found with $c = \pm1$, such that
$x \mp y = 0$.  In this case, the geodesics are coincident at the
starting intersection and we treat this as though the geodesics are {\it
nearly} coincident.  Therefore, there are intersections at $\mathbf T' =
[s, \pm s]$ where $s$ is the displacement to successive conjugate points
along one of the geodesics given by the condition $m_{12} = 0$.
\item
Otherwise, canonicalize the intersection by minimizing
the $L_\infty$~distance to $\mathbf 0$ as we did for finding the closest
intersection.
\end{enumerate}

Similar strategies can be used when finding all the intersections.  For
details, see the implementation.

\section{Implementation}\label{implementation}

The methods described in this paper are included in version 2.3 of
GeographicLib \citep{geographiclib23}.  The core methods are provided by
the C++ class {\tt Intersect}.  This can utilize the solution of the
geodesic problem in terms of Taylor series (suitable for $\abs
f \le \frac 1{50}$) or elliptic integrals (suitable for any value of
$f$).
However, remember that the testing for these methods has been limited to
$\abs n \le 0.12$; so this class should not be used for more eccentric
ellipsoids.  The library also includes a utility, {\tt IntersectTool},
that
allows intersections to be computed on the command line.

The routines are optimized by the use of the {\tt GeodesicLine} class.
This describes a geodesic specified by a starting point and azimuth.
Once this has been set up, computing positions at an arbitrary
displacement along the line is about two times faster than solving the
direct geodesic problem afresh.  The major computational cost is then
the solution of the inverse problem in Step 2 of the basic algorithm.

The most useful capabilities are probably finding the closest
intersection for two geodesics and finding the intersection of two
geodesic segments.

Finding the solution for the closest intersection for randomly chosen
geodesics (uniformly distributed positions and azimuths) on the WGS84
ellipsoid, $f = 1/298.257223563$, requires on average $3.16$ solutions of
the inverse geodesic problem.  The first solution found $\mathbf T_0
= \mathbf b(\mathbf 0)$ is accepted $99.55\%$ of the time in Step 2(b).
$\mathbf T_0$ is {\it eventually} accepted an additional $0.38\%$ of the
time, with $\mathbf T_i$ for $i > 0$ accepted the remaining $0.07\%$ of
the time.  If $\mathbf T_0$ is not accepted right away an average of
$1.26$ additional applications of the basic algorithm are required.  The
mean number of applications of the basic algorithm is $1.0056$.

These results are essentially unchanged for finding the intersection of
geodesic segments because, for the WGS84 ellipsoid, the mean number of
times the basic algorithm needs to be evaluated at a corner point is
minuscule: $1.5\times f^5 \sim 10^{-12}$.  So the cost is then merely
that of a closest intersection calculation plus the cost of the two
inverse geodesic calculations needed to determine the azimuths for the
segments.  On a $3.4\,\mathrm{GHz}$ Intel processor, the average time to
compute the intersection of two segments on the WGS84 ellipsoid is about
$12\,\mu\mathrm s$.  The intersection point can be found either by
computing the point a displacement $x$ along geodesic $X$ or the point a
displacement $y$ along $Y$.  Due to numerical errors, these two computed
points do not, in general, coincide; however, they are within about
$0.025\,\mu\mathrm{m}$ of each other and this is a useful gauge of the
overall accuracy of the implementation.  The errors in the displacements
$x$ and $y$ may be larger than this if the geodesics intersect at a
small angle, increasing by about the cosecant of that angle.

\section{Discussion}

We have given algorithms that reliably find the intersections of
geodesics.  Our contributions include an important correction to the
basic method provided by {\it BML} followed by a systematic application
of this method with selected starting points to obtain the closest
intersection and the intersection of geodesic segments.

This allows operations such as computing the union and intersection of
polygonal shapes on the surface of an ellipsoid.  Coincident geodesics
are treated correctly; this is important, for example, in analyzing two
polygonal regions that share a boundary.

We offer a conjecture that whenever two geodesic segments intersect, the
intersection is the one that is closest to the midpoints of segments.

For terrestrial ellipsoids, the basic method returns the closest
intersection in most cases.  However, taking the extra steps to ensure
that the closest intersection is always returned is crucial: it relieves
the user of the library from handling possibly erroneous results and the
runtime cost is minimal.

Our treatment applies to moderately eccentric oblate and prolate
ellipsoids, $\frac45 \le b/a \le \frac54$; we expect similar techniques
to apply to more complex surfaces, e.g., a triaxial ellipsoid.

{\it BML} also consider the interception problem, namely finding the
point on a geodesic closest to some given point.  I touch on some
aspects of this problem in Appendix~\ref{interceptsec}.

\section{Data Availability}

The source code for the C++ implementation of the intersection routines
is available
at \url{https://github.com/geographiclib/geographiclib/tree/r2.3}.
This is documented in \citet{geographiclib23}.

\ifeprint \bibliography{geod} \fi 

\appendix

\section{The basic algorithm}\label{spherical}

\begin{figure}[tbp]
\begin{center}
\includegraphics[scale=0.75]{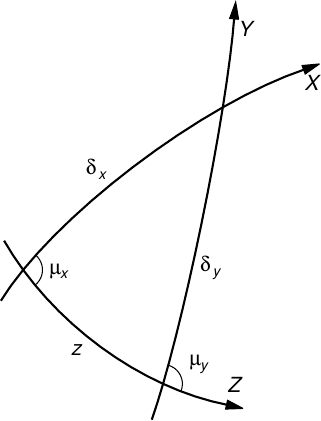}
\end{center}
\caption{\label{triangle}
The spherical trigonometry problem for the basic algorithm.}
\end{figure}%
The core of the algorithm given by {\it BML} starts with a tentative
intersection $\mathbf S = [x, y]$ and produces an improved intersection
$\mathbf S' = [x', y']$.  The steps are as follows:
\begin{enumerate}
\item
Solve the direct geodesic problem for each geodesic to determine
$(\phi_x, \lambda_x)$, $\alpha_x$ and $(\phi_y, \lambda_y)$, $\alpha_y$
at the tentative intersection.
\item
The geodesic $Z$ connects $(\phi_x, \lambda_x)$ and
$(\phi_y, \lambda_y)$.  Solve the corresponding inverse geodesic
problem and find the distance $z$ and the azimuths at the endpoints
$\gamma_x$ and $\gamma_y$.  (We use $\gamma$ to denote the
azimuths on geodesic $Z$.)
\item
If $z = 0$, set $[\delta_x, \delta_y] = \mathbf 0$ and go to the last step.
\item
The geodesics $X$, $Y$, and $Z$ form an ellipsoidal triangle with one
edge of length $z$, and adjacent angles $\mu_x$ and $\pi-\mu_y$ where
$\mu_x = \gamma_x-\alpha_x$ and $\mu_y = \gamma_y-\alpha_y$; see
Fig.~\ref{triangle}.  The intersection is found by determining the
lengths of the other two sides; however, if the triangle is inverted,
i.e., $\mu_y-\mu_x \pmod{2\pi} < 0$, first change the signs of $\mu_x$
and $\mu_y$.
\item
An approximate solution to the ellipsoidal problem is found by solving
the corresponding problem on a sphere of radius $R$; thus lengths are
converted to central angles, for example, by $\zeta=z/R$.  The two sides
are then given by
\begin{align}
\frac{\delta_x}R &= \atanx
{ \sin \mu_y \sin \zeta }
{ \sin \mu_y\cos \mu_x\cos \zeta
  - \cos \mu_y\sin \mu_x },\label{spherex}\displaybreak[0]\\
\frac{\delta_y}R &= \atanx
{ \sin \mu_x \sin \zeta}
{ -\sin \mu_x\cos \mu_y\cos \zeta
 + \cos \mu_x\sin \mu_y }.\label{spherey}
\end{align}
\item
Set $\mathbf S' = \mathbf S + [\delta_x, \delta_y]$ as the improved
intersection.
\end{enumerate}
These steps define a mapping $\mathbf S' = \mathbf h(\mathbf S)$ and
this mapping can be repeated to give $\mathbf S'' = \mathbf h(\mathbf
S') = \mathbf h^2(\mathbf S)$, $\mathbf S''' = \mathbf h^3(\mathbf
S)$, etc.  In the limit, we have
\begin{equation}\label{g-def}
\mathbf b(\mathbf S) = \lim_{n\rightarrow\infty} \mathbf h^n(\mathbf S).
\end{equation}
This mapping $\mathbf b(\mathbf S)$ constitutes the ``basic algorithm''.
{\it BML} apply this starting with the points defining the geodesics,
i.e., $\mathbf S = \mathbf 0$, computing $\mathbf b(\mathbf 0)$ as the
intersection.

A few points require explanation:
\begin{itemize}
\item
Equations~(\ref{spherex}) and (\ref{spherey}), which solve the spherical
triangle, are obtained from the cotangent rules given
by \citet[\S44]{todhunter86}.  Superficially they are the same as
Eqs.~(3) and (4) in {\it BML}, except that here we ensure that the
correct quadrant is chosen for $\delta_x/R$ and $\delta_y/R$; we use the
heavy ratio lines in these equations to mean that the quadrant for the
arctangent is given by the {\it separate} signs of the numerator and
denominator of the fraction; this functionality is provided by the {\tt
atan2} function in many computer languages.  This guarantees that
$\delta_x$ and $\delta_y$ are
consistent; thus, on the sphere, they advance to the same intersection,
and, provided that $\zeta < \pi$, this intersection is the closest.
\item
Equations (\ref{spherex}) and (\ref{spherey}) are well defined if {\it
either} the numerator or the denominator of the right-hand sides
vanishes.  However, the equations are indeterminate if {\it both} the
numerator and the denominator vanish, i.e., when
$\sin \mu_x \approx \sin \mu_y \approx 0$.  This corresponds to {\it
coincident} geodesics $X$ and $Y$.  In this
case, a suitable solution is $\delta_x = \frac12 z \cos \mu_x$ and
$\delta_y = -\frac12 z \cos \mu_y$ which places the intersection midway
between the starting points; also set a ``coincidence flag'', $c = \pm1
= \cos\mu_x\cos\mu_y$, to indicate whether the geodesics are parallel or
antiparallel.  Another case of coincidence is if $\zeta\approx0$ and
$\abs{\sin\mu_x \mp \sin\mu_y} \approx \abs{\cos\mu_x \mp \cos\mu_y} \approx
0$ (i.e., $\mu_y - \mu_x \approx 0$ or $\pi$), in which case set the flag
$c = \pm1$.  If the geodesics are not coincident, set $c = 0$.  A
suitable tolerance for the approximate equality here $3 \epsilon$, where
$\epsilon$ is the precision of the floating-point number system; for
standard double-precision arithmetic, we have $\epsilon = 2^{-52}$.
\item
The algorithm always converges to an intersection, i.e., the limit in
Eq.~(\ref{g-def}) exists, and the convergence is quadratic.  As the
algorithm progresses, the sides of the ellipsoidal triangle become
shorter and, in the limit, the spherical solution degenerates to the
solution for a plane triangle; as a consequence, the quadratic
convergence does not depend on the value picked for $R$.  A suitable
convergence criterion is $\abs{[\delta_x, \delta_y]}
< \epsilon^{3/4} \pi R$.  When this inequality is satisfied, the errors
in the updated intersections are, as a result of the quadratic
convergence, less than $\epsilon^{3/2}\pi R$, well below the roundoff
threshold.  Typically about 3 iterations are required for convergence.
\end{itemize}

\section{Geodesic interceptions}\label{interceptsec}

{\it BML} also investigate the geodesic interception problem, finding
the minimum distance from a point $P$ to a geodesic $AB$.  I give here
two improvements to their algorithm.  I adopt the notation used by {\it
BML}, who treat $A$ as the tentative interception and compute $X$, a
point on $AB$ as an improved interception.  Equation (10) in {\it BML}
can be replaced by
\begin{equation}\label{intercept}
  \frac{s_{AX}}R =\atanx{ \sin(s_{AP}/R)\cos A}{\cos(s_{AP}/R)},
\end{equation}
with the heavy ratio line indicating (as above) that the quadrant is
given by the signs of the numerator and denominator, using, for example,
the {\tt atan2} function;
$A$ is then replaced by $X$ and the process is repeated.  A similar
formula is given by \citet{mllario21}, with the caution that it
shouldn't be used for $s_{AP}/R \rightarrow \frac12\pi$.  In fact, this
formula is well-behaved in this limit (returning $s_{AX}/R
= \pm \frac12\pi$ depending on the sign of $\cos A$).  Repeated
application of Eq.~(\ref{intercept}) converges to $\cos A \rightarrow
0$, the condition that the distance from $P$ to the geodesic is an
extremum; but, because this equation does not fully capture ellipsoidal
effects, the convergence is slower than quadratic.  This problem can be
corrected using Eq.~(\ref{intercept}) for the first iteration and
\begin{equation}\label{intercept2}
  \frac{s_{AX}}R = \frac{m_{AP}}R
  \frac{\cos A}{(m_{AP}/s_{AP})\cos^2 A + M_{AP} \sin^2 A}
\end{equation}
for subsequent ones.  Here $m_{AP}$ and $M_{AP}$ are the reduced length
and geodesic scale for the geodesic connecting $A$ and $P$, which can be
inexpensively computed as part of the geodesic inverse
problem \citep[\S3]{karney13}.

This addresses the core interception problem (the equivalent of the
basic algorithm for the intersection problem).  Seeking a global optimum
solution is bedeviled by the fact that, as shown by \citet{botnev15},
successive minima in the distance can be arbitrarily close.  This is a
sharp contrast with the intersection problem where there is a minimum
distance between intersections.  This problem is addressed
by \citet{botnev19}.

\ifeprint\else 
\bibliography{geod}
\fi 

\end{document}